\documentclass[conference]{IEEEtran}
\IEEEoverridecommandlockouts
\usepackage[noadjust]{cite}
\usepackage{amsmath,amssymb,amsfonts}
\usepackage{algorithmic}
\usepackage{graphicx}
\usepackage{textcomp}
\usepackage{xcolor}
\usepackage[capitalise]{cleveref}
\usepackage{siunitx}
\usepackage{subcaption}

\DeclareSIUnit\operations{op}
\usepackage[colorinlistoftodos,prependcaption,textsize=tiny]{todonotes}

\def\BibTeX{{\rm B\kern-.05em{\sc i\kern-.025em b}\kern-.08em
    T\kern-.1667em\lower.7ex\hbox{E}\kern-.125emX}}
\begin{document}

\title{Detecting GNSS misbehavior leveraging secure heterogeneous time sources}

\author{\IEEEauthorblockN{Marco Spanghero}
	\IEEEauthorblockA{\textit{Networked System Security Lab} \\
		\textit{KTH Royal Institute of Technology}\\
		Stockholm, Sweden \\
	marcosp@kth.se}
	\and
	\IEEEauthorblockN{Panos Papadimitratos}
	\IEEEauthorblockA{\textit{Networked System Security Lab} \\
		\textit{KTH Royal Institute of Technology}\\
		Stockholm, Sweden \\
	papadim@kth.se}
}

\maketitle

\begin{abstract}
	Civilian Global Navigation Satellite Systems (GNSS) vulnerabilities are a threat to a wide gamut of critical systems. 
 GNSS receivers, as part of the encompassing platform, can leverage external information to detect GNSS attacks. Specifically, cross-checking the time produced by the GNSS receiver against multiple trusted time sources can provide robust and assured PNT. In this work, we explore the combination of secure remote, network-based time providers and local precision oscillators. This multi-layered defense mechanism detects GNSS attacks that induce even small time offsets, including attacks mounted in cold start. Our system does not require any modification to the current structure of the GNSS receiver, it is agnostic to the satellite constellation and the attacker type. This makes time-based data validation of GNSS information compatible with existing receivers and readily deployable.
\end{abstract}

\section{Introduction}

Global Navigation Satellite Systems (GNSS) provide ubiquitous time and navigation. On the other hand, civilian GNSS vulnerabilities have been recently increasingly visible (\cite{Thombre2018, psiaki2016gnss}).  The vast majority of civilian GNSS receivers (currently) rely on signals without security features (e.g. cryptographical protection at the physical layer or of the navigation messages \cite{Fernandez-Hernandez2016,anderson2017chips}). There are proposals for cryptographic protection for civilian GNSS (\cite{Gamba2021ComputationalPlatforms,Cucchi2021AssessingReceiver,Motella2020AReceiver}), such as the Galileo Open Service Navigation Message Authentication and the upcoming GPS Chimera, however, operational deployment requires time \cite{ Gotzelmann2021GalileoProvision}. Furthermore, cryptographic methods are likely to be implemented in the new generation of receivers (as they require changes to the receiver structure), not covering receivers already deployed in the field. Similarly, lack of low-level measurements, high computational requirements, or hardware support, somewhat limits the applicability of defenses based on signal properties (e.g., angle of arrival). In addition, adversarial control of the GNSS receiver Position, Navigation, and Time solution (PNT) is still possible, even when authenticated signals are considered: cryptographic protection cannot fully address replay/relay attacks (\cite{Lenhart2022, Seco-Granados2021}).

A key observation is that GNSS receivers are often integrated into devices that provide network connectivity and computational power: these can be used to validate the GNSS-provided PNT by comparing the GNSS-obtained information with other reference sources. 
In this work, we consider time reference sources: leveraging the onboard clock (\cite{papadimMilcom2008, jafarnia2013PNT,Arafin2016DetectingOscillators,Hwang2014, Spanghero2022}) or external time information (\cite{spangheroGNSS20, kzmsppPLANS2020}). 
As long as the attack on the GNSS receiver causes the time part of the PNT to deviate from the actual time, the attack can be detected effectively.
The challenge is to detect sophisticated adversaries aiming at spoofing the GNSS receiver while, possibly, controlling remote time providers, in a coordinated manner. Such adversaries evade existing time-based defense mechanisms. Insecure remote time references can be tampered with by leveraging known vulnerabilities, hence trustworthy reference time information must be provided, practically resorting to secure time transfer methods. 

Using local timekeeping is as effective as the on-board clock hardware. Moreover, local time-based tests cannot thwart attacks at \textit{cold start}. This can be achieved by obtaining time information from external, networked servers, as long as those can be authenticated, to prevent networked-based attacks combined with GNSS attacks. Nonetheless, intermittent network connectivity can deprive the receiver of such external time sources. Naturally, one can combine the two ideas, GNSS time (and implicitly the overall PNT) validation based on recurring interaction with networked time servers and, in the meantime, validation based on the local clock hardware.

In this work, we propose exactly this: a method to seamlessly detect spoofing attacks on the GNSS receiver, based on the consistency of the time information from different time providers. Our approach relies on the progressive refinement of the control time reference to optimize the detection threshold, the needed computational power, and energy consumption. Our combination of cryptographically secure coarse time references (Roughtime), online secure time servers (NTS), and an accurate local, on-board clock ensemble provides reliable GNSS attack detection. Heterogeneous clock sources from multiple providers are combined based on availability and required security or accuracy levels.

The rest of this paper is organized as follows: \cref{section:background} presents the relevant related work, \cref{section:sys-adversary-model} analyzes the assumptions made regarding the nature of the adversary and the system we consider here, \cref{section:methodology,section:implementation} discuss the detection method developed and the experimental setup, \cref{section:results-conclusions} analyzes the spoofing detection results, \cref{section:conclusion} concludes the work and discusses possible extensions to the presented method. 

\section{Related Work}
\label{section:background}

First, we review GNSS attacks targeting (or affecting to a certain extent) the time of the GNSS PNT solution and relevant countermeasures. Then, we analyze countermeasures that leverage additional providers of time or position and navigation. 

\textbf{Attacking GNSS receivers.} GNSS receivers are the target of multiple types of attacks with successful overtake of devices in real-life operational scenarios  (\cite{tippenhauer2011requirements, Kerns2014, Bhatti2017, Ioannides2016}). Spoofing attacks can be carried out either by simulation, co-simulation using a receiver spoofer, or meaconing/replay. Simulation attacks, albeit effective in altering the PNT solution of the victim GNSS receiver, do not maintain any synchronization with the original signal. Low-cost software-defined radio (SDR) hardware and software tools availability made spoofing by simulation accessible to the masses, increasing the risk (\cite{KexiongAllBelongToUs2018, Feng2021, HuangL2015}) even in multi-constellation \cite{LeksellTGalileo2021} and multi-frequency scenarios \cite{SDRMultiFrequency2018}.

More advanced methods, based on knowledge of the state of the target spoofing constellation, require the adversary to operate a GNSS receiver practically building an intermediate spoofer. Implementations of effective receiver-spoofer matched adversaries exist and are capable of advanced signal lift-off of the victim receiver (\cite{HumphreysAssessingSpoofer, Maier2018}). Attacks implementing lift-off require careful synchronization of code and Doppler shift and mobility of the phase center of the victim antenna, making deployment targeting mobile adversaries complex. On the other hand, examples targeting static timing-dedicated receivers exist, showing that such attacks effectively target smoothly the time solution at the GNSS receiver, specifically when applied to phasor measurement units in smart grids (\cite{Shepard2012c, Humphreys2012, Jiang2013, Zhu2016}). As a specific case of fully synchronized adversaries, Time Synchronization Attacks (TSA) target the time solution of the GNSS receiver, minimally disturbing the location or navigation part \cite{Zhang2013}.

Signal replaying/relaying, also known as meaconing, allows any attacker to replay/relay signals corresponding to a different place or time to a victim receiver, effectively shifting the PNT solution \cite{Lenhart2022}. Cryptographically secure signals can also be the target of advanced replay-based attackers, despite the additional security features. Specifically, techniques that use Secure Code Estimation and Replay (SCER) can be effective in spoofing even cryptographically-hardened receivers (\cite{humphreys2013detection, Arizabaleta2019, Gallardo2020}).

\textbf{Protecting GNSS receivers.} To target the diffusion of attack vectors, countermeasures target various levels of the receiver implementation, aiming to improve the robustness of the PNT. At the signal level, Carrier-over-Noise ($C/N_0$) measurements provide direct insight into the GNSS signal strength at the receiver front-end. Combination of the $C/N_0$ with the receiver front-end gain allows for the detection of spoofing signals by correlation of the received power variation or distortion of the power envelope (\cite{Akos2012, papadimMilcom2008, Lo2019, wesson2017gnss}). 

Techniques using the Doppler shift of the received signals (commonly known as Doppler test) allow the detection of spoofed signals based on the transmitter frequency error (\cite{papadimMilcom2008, psiaki2013antenna}). Other techniques, involving the deformation or splitting of the correlation peaks during spoofing, also allow the GNSS receiver to detect or exclude specific GNSS signals originating from an adversary, in particular when combined with Receiver Autonomous Integrity Monitoring (RAIM) techniques (\cite{Jada2021, Sathaye2020}). 

Most of these methods require the receiver to expose low-level signal features and raw measurement information. Although this is an increasingly common feature, especially in mobile phones \cite{Miralles2018}, processing this information requires dedicated algorithms and often considerable computational power. 
Additionally, techniques requiring modifications to the acquisition and tracking channel structure in the GNSS receiver are complex to deploy in the current receiver generation.

\textbf{Data level GNSS validation.} The information of the PNT solution can directly be compared with other providers of time or navigation to detect inconsistencies. These methods require low computation and limited connectivity and can be highly modular to protect different aspects of the PNT solution. Solutions targeting the navigation part of the PNT often rely on external sensors (e.g., IMUs) to compare acceleration and velocity obtained from the GNSS receiver (\cite{Curran2017OnTU, Broumandan2018, Clemens2022}). On the other hand, detection in mobile devices is often limited due to the poor performances of low-tier inertial sensors (i.e. rapid growing integral errors due to drift and/or bias in accelerometers and gyroscopes). The combination of sensors and localization with respect to terrestrial networks is investigated in \cite{LiuWPP:2023}.

Validation of the GNSS receiver provided time is also possible using alternative time providers. Solutions considering single or multiple precision embedded clocks proved successful in detecting offsets and drift in the time solution due to an adversary (\cite{Arafin2016DetectingOscillators, Arafin2017, Spanghero2022, Hwang2014}). Additionally, leveraging network time allows even sparsely connected receivers to validate the accuracy of the GNSS-provided time against selected remote references (\cite{kzmsppPLANS2020, papadimMilcom2008, ODriscol2020}). A combination of these methods is still partially unexplored in the context of validating the GNSS time solution. 

Advanced attackers targeting network time providers can control individual time servers or entire pools (\cite{perry2021, Deutsch2018, Malhotra2017}). Methods to improve the security of the Network Time Protocol (NTP) structure leverage cryptographical extensions to the standard but initial adaptations proved to be not secure \cite{rfc5906}, due to vulnerabilities in the implemented algorithms. The Internet Engineering Task Force (IETF) standardized a new security extension for NTP, Network Time Security (NTS), aiming principally at secure time distribution in networked systems \cite{ietf-ntp-using-nts-for-ntp-28}. The most recent evolution of the standard addresses most of these issues, making NTS an important candidate for secure network time transfer. 

Additionally, Google Roughtime achieves a high level of time distribution assurance by providing digitally signed and non-repudiable coarse time information \cite{ietf-ntp-roughtime-07}. Due to the strong limitations of Roughtime accuracy and the higher computational requirements, its application in GNSS time validation is limited. Nevertheless, the time validation at cold-start perfectly fits Roughtime, particularly when combined with more accurate methods like NTS.

\section{System and Adversary model}
\label{section:sys-adversary-model}
We consider a commercially available, off-the-shelf GNSS receiver connected to a system capable of providing the required computational power and connectivity. Additionally, we assume the system is provisioned with one or more on-board precision reference oscillators that can be used to provide high-quality, stable reference time.

The aim of the GNSS receiver is to solve \cref{eq:PNT_eqn_good}, where $p$ is a $n \mathrm{x} 1$ vector of pseudoranges observations, $H$ is an $n\mathrm{x}n$ observation matrix, $x=[x,y,z,t]$ is the receiver state vector of location and time offset at the victim receiver, and $v$ is the total system noise.

\begin{equation}
    p = Hx + v
    \label{eq:PNT_eqn_good}
\end{equation}

All GNSS systems rely on precise timekeeping to measure the distance between the receiver and each satellite. In \cref{eq:pseudorange}, the satellite-receiver pseudorange for satellite (s) in the $i$-th band is a function of the receiver's own best guess of the time at signal reception time, $\bar{t}_r$.
\begin{equation}
    P^{(s)}_{r,i} = c(\bar{t}_r - \bar{t}^{(s)})
    \label{eq:pseudorange}
\end{equation}

We can rewrite the previous expression as a function of the geometrical range $\rho_r^{(s)}$, the receiver and satellite clock biases $dt$ and $dT$ respectively, referenced to the respective timescales. To simplify the discussion, we purposely exclude any ionospheric and tropospheric correction terms as they only represent additive errors. The resulting equation is shown in \cref{eq:pseudorange_geometerical}, where $\epsilon_p$ represents all the satellite, measurement, and instrumentation delays in the system.

\begin{equation}
    P^{(s)}_{r,i} = \rho_r^{(s)} + c(dt_r(t_r) + dT^{(s)}(t^{(s)})) + \epsilon_p
    \label{eq:pseudorange_geometerical}
\end{equation}

In \cref{eq:pseudorange_geometerical}, the transmission time at the satellite can be obtained based on Time of Week (TOW) information distributed by each satellite, while the receiver time offset (which, practically, is the time difference between the receiver's time scale and the constellation time scale) is obtained at first with a reasonable guess and progressively refined. This is possible once the tracking loops locks-in the satellite signals and the receiver decodes the navigation message words. The principle of common reception time is often used in commercial receivers and allows the GNSS receiver to:
\begin{itemize}
    \item Reference all measurements to the earliest arriving subframe, and assign an arbitrary travel time, based on the estimated distance of the orbital plane.
    \item Obtain the individual (relative) pseudoranges by adding the early-late delay of each channel to the reference
\end{itemize}
Based on the information obtained from the Locked Delay Loop and its own reference clock, the GNSS receiver can obtain the actual pseudorange for the reference signal and derive all the other ones and can calculate a full PNT solution. 

The time (or position) of the GNSS receiver can be controlled by an external adversary by carefully crafting valid GNSS signals, consistent with the intention of the adversary. Due to the open structure of GNSS signals, adversaries can create signals that are valid from modulation, information content, and spectral component perspectives.

Notably, we are largely agnostic in terms of the specific type of attack mounted by the adversary to control the victim receiver. We consider a generic attacker model that causes variations, either abrupt or smooth, in the receiver time solution (independently of the adversarial action target, which could be the receiver's time itself or its position). To the best of our knowledge, it is unfeasible for a GNSS attacker to avoid perturbing the time solution in order to achieve a successful position overtake. If achieved, any such attack would be undetectable by any countermeasure testing the receiver clock solution, including our own. On the contrary, overtaking the time solution without large modifications of the navigation solution is feasible in the case of Time Synchronization Attacks (TSA). 

Without loss of generality, we consider the case of a single constellation GNSS receiver, specifically GPS: this is not a limitation to the applicability of the method presented as it can be extended to be compatible with different GNSS systems or even combinations of multiple constellations (i.e. multi-constellation GPS and Galileo receivers).
More in detail, let's consider the modified model from \cref{eq:PNT_eqn_good} for a GNSS receiver under adversarial control, where $f$ is the component of adversarial actions:
\begin{equation}
    p = Hx + f + v
    \label{eq:PNT_eqn}
\end{equation}
A simplistic adversary would generate signals that correspond to a specific $x$ solution and transmit them to the victim receiver, which would observe a superposition of real and adversarial signals. Given enough power advantage, the adversary could capture the victim receiver, as the interference between the real and fake signals would initially cause a loss of lock in the victim similar to a jamming event. At the following signal re-acquisition, the higher-power adversarial signals would be acquired instead of the real ones, causing the GNSS receiver to obtain a fake PNT solution. Due to the lack of synchronization between the GNSS frames in the real and fake signals, sharp discontinuities in the PNT solution can be observed in the time component.

A more sophisticated adversary can instead construct signals that not only have the same structure as the real ones but also are aligned in the code phase and Doppler shift domains to the original signals. In practice, this allows the attacker to capture the tracking loop at the victim receiver without causing a loss of lock and to smoothly change the receiver state vector. These advanced attackers are required to have knowledge of the current constellation status at the victim receiver and its mobility. Additionally, they must track the victim for the duration of the attack to maintain control of it. Due to the high degree of similarity with the original signal at the onset of the attack, they are often more subtle and difficult to detect due to the slow variations introduced in the PNT solution.

Other complex attacks that can be mounted against cryptographically enhanced receivers (such as Secure Code Estimation and Replay  (SCER)) are beyond the scope of this work. We believe that the likelihood of such attacks being deployed in real-life scenarios against low-value targets is low given the sophistication and knowledge required for the adversary to be successful. Nevertheless, it is unfeasible for an attacker to maintain perfect alignment of both the time and position solution at the onset of a SCER attack (\cite{psiaki2013}), hence considering SCER attackers is worth a dedicated investigation.

Extending the attacker's capabilities, its influence might not be limited to the GNSS receiver. An adversary can overtake network-provided references in a coordinated manner to attempt to minimize the GNSS tempering detection probability. Such an adversary can control, deny, limit, or tamper with the victim's access to the network-provided time. Attacks on the Network Time Protocol (NTP) show that it is possible for a strategically placed attacker to modify the perceived time at the victim by fully or partially controlling the victim's interaction with the remote NTP time server (or server pool). Attackers can replay, modify or delay any unprotected packet that is used to obtain synchronization between the victim and the remote server. Additionally, impersonation of time servers is also possible, which would allow the attacker to have full control of the victim's time solution.

\section{Methodology}
\label{section:methodology}
Due to the heterogeneity of time sources available to the system (which can consist of either enhancements to the system components, or provided by network-connected entities, beyond the GNSS receiver), clear assumptions need to be made regarding the level of trustworthiness and performance required. We consider any reference that is contained within the hardware boundaries of the GNSS-enabled system (e.g. its packaging) to be trusted, meaning the hardware our solution is implemented on cannot be compromised. Additionally, we assume that the GNSS receiver (when not under adversarial control) is always the most accurate PNT source available to the system. To guarantee the best level of service, the system always uses the most accurate PNT available, unless a discrepancy against any reference source is detected and the PNT is deemed to be under attack. In that case, the system will choose the most trusted time reference.

On the contrary, we consider any reference time source external to the GNSS-enabled system (e.g., network-provided time) to be not trusted unless cryptographically protected, corroborating the validity of the provided time information and authenticating communication. Nevertheless, we do not exclude the possibility for the system to use both trusted and untrusted network time providers, but we assume that in case of a discrepancy among the time solutions, the system will resort to the most trusted provider, even at a penalty of reduced accuracy. In addition, we assume that access to the network might be unavailable (due to a benign fault or adversarial action), hence we do not require a constant exchange of information with the connected time references. Moreover, the receiver can be either in cold start at the beginning of the attack (practically, this means the receiver has no current knowledge of the state of the constellation and the time offset of its embedded oscillator) or the receiver has already acquired a valid solution (and obtained recent constellation status updates). 

\begin{figure}[h!]
    \centering
    \includegraphics[width=\linewidth]{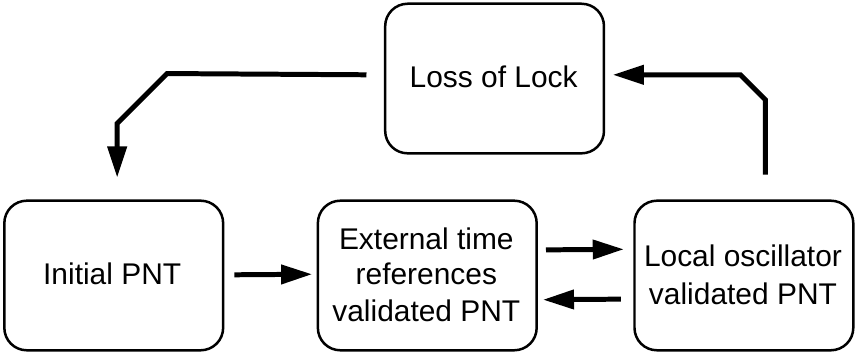}
    \caption{Event-based switching for data-level GNSS time validation.}
    \label{fig:switching-schema}
\end{figure}

\subsection{Time-based defenses through progressive refinement}

Intuitively, the challenge is to progressively refine the level of confidence the system has regarding the GNSS time solution (and by extension the other components of the PNT). This can be achieved by "sifting" the GNSS-provided time using progressively more precise time sources to achieve continuous validation and tracking of the receiver clock offset. More importantly, each component is part of a scalable and extensible method: it is possible to combine them to obtain an all-around solution.  The receiver state and connectivity availability events trigger changes for the countermeasure state machine (\cref{fig:switching-schema}). The initial PNT solution time can be validated using a (possibly, coarse if needed) component (\cref{sec:roughtime_comp}), and the system belief of the GNSS clock correctness can be progressively refined using a combination of more accurate network time sources (\cref{sec:nts_comp}) and local oscillators (\cref{sec:local_comp}), which provide high-resolution short-term information even without an active internet connection. We discuss next the individual components targeting different aspects of the time validation problem and how to combine them (\cref{sec:combining}) to achieve assurance of the time solution at the GNSS receiver.

\subsection{Component 1 - Roughtime}
\label{sec:roughtime_comp}
Given an adversary could be present and an attack ongoing before the receiver is powered on, it is important to protect the initial acquisition process, bounding the initial PNT solution (and by extension the initial receiver offset) within an error of a trusted time source. The first component we present targets exactly this problem: how to validate in a trustworthy manned the first update of the PNT solution once the receiver reports a successful lock. Google Roughtime provides secure and digitally signed time information from an ecosystem of trusted providers. To obtain secure time verification, a client creates a request to a remote time reference and obtains a digitally signed reply containing the authenticated time. 
Once the receiver solves \cref{eq:PNT_eqn} for $dt_r$, it obtains the time offset between the receiver counter clock and the GPS timescale. The PNT solution is often expressed as aligned to the UTC scale, which means that the receiver also compensates for leap seconds that correspond to the offset between the GPS timescale and UTC. Once a UTC-aligned PNT solution is obtained, our detection system can verify if the UTC timestamp is aligned to a trusted UTC timescale such as the one provided by Roughtime.

A Roughtime measurement is $T^{(s)}_{RT} = \{t^{(s)}, R^{(s)}\}$, which are the absolute timestamp of UTC time and the server's confidence radius. The confidence radius is provided as Roughtime is designed to provide coarse time. Consider the following $t_{GNSS}, t^{(s)}_{RT}$, the receiver provided UTC time and the Roughtime timestamp. We can write the following binary hypothesis test for any Roughtime server available in the ecosystem:

\begin{equation}
    \mathcal{H}_i =
    \begin{cases}
        \mathcal{H}_0\ \mathrm{if}\ |t_{GNSS} - t^{(s)}_{RT}| < R^{(s)} \\
        \mathcal{H}_1\ \mathrm{otherwise}
    \end{cases}
    \label{eq:hypothesis-rt}
\end{equation}

The outcome of the test in \cref{eq:hypothesis-rt} determines if the GNSS-provided UTC scale is aligned with a trusted UTC scale.

Once the initial time update is validated, the system is able to anchor (at least coarsely) the local time scale with the trusted remote one. Nevertheless, as GNSS provided time is orders of magnitude more accurate than the Roughtime-provided one, progressive refinement of the time solution is required. Two strategies are available, based on the available network connectivity. 

\subsection{Component 2 - Network time security}
\label{sec:nts_comp}

If connectivity is available throughout the operation of the receiver, it is possible to monitor the GNSS time solution based on one (or multiple) trusted network time servers. Compared to the Roughtime-based verification, which is cryptographically expensive and coarse only, network time servers provide lightweight, secure time but require a key establishment protocol, building on Transport Layer Security (TLS) - with an extension to Network Time Protocol (NTP), the Network Time Security (NTS) as specified in IETF RFC8915 \cite{ietf-ntp-using-nts-for-ntp-28}.

Direct verification of the GNSS time offset is possible using the NTS server pools, by checking the PNT timescale against the remote time server. Intuitively, two ways to achieve this exist. First, the system can check at each PNT update if the GNSS obtained time is consistent with the NTS provided time. This is performed by simply differencing the GNSS-obtained time and the one (or multiple) obtained from the NTS pool, with an approach similar to \cref{eq:hypothesis-rt}. The modified hypothesis test, valid for each NTS server reachable at any epoch, is shown in \cref{eq:hypothesis-nts}, where $t^{(s)}_{NTS}$ is the NTS derived time and $\lambda_{T}$ is a threshold obtained based on the quality of the remote NTS source (which can be configured based on the application requirements).

\begin{equation}
    \mathcal{H}_i =
    \begin{cases}
        \mathcal{H}_0\ \mathrm{if}\ |t_{GNSS} - t^{(s)}_{NTS}| < \lambda_{T}\\
        \mathcal{H}_1\ \mathrm{otherwise}
    \end{cases}
    \label{eq:hypothesis-nts}
\end{equation}

Alternatively, variations in the frequency offset between the NTS server and the GNSS receiver time can also be used to identify an attacker, but investigation of other key performance indicators in NTS is useful to track the validity of the GNSS time is left for future work, as this work focuses on techniques compatible both with NTS and Roughtime.

\subsection{Component 3 - Local oscillator ensemble}
\label{sec:local_comp}

Constant connectivity is not required if the system is provided with one (or multiple) local stable clocks. Precision oscillators are useful in continuously tracking and monitoring the state of the receiver clock offset. The local clock can be used to establish a local timescale whose initial offset to the GNSS timescale is known and potentially zero. Intuitively, within the provided stability window of the local oscillator, the progression of time in the GNSS receiver and in the local timescale is identical. During an attack causing the manipulation of the GNSS receiver time offset, a discrepancy is measured between the local timescale and the GNSS-provided one.

Consider the model from \cref{eq:clock_model_gnss}, normally used for a GNSS clock, where $d,b$ are the clock drift and bias respectively and $w_b, w_b$ are the process noise values for clock drift and bias.
\begin{equation}
    \begin{pmatrix}
        \dot b \\ \dot d
    \end{pmatrix}
    =
    \begin{pmatrix}
        0 & 1 \\ 0 & 0
    \end{pmatrix}
    \begin{pmatrix}
        b \\ d
    \end{pmatrix}
    +
    \begin{pmatrix}
        w_b \\ w_d
    \end{pmatrix}
    \label{eq:clock_model_gnss}
\end{equation}

For the problem at hand, which is verifying the consistency of the GNSS timescale, we can write the inter-scale clock bias $\Delta_{1,2}$ between the local oscillator timescale and the GNSS one as an uncorrelated random-walk variable. We can then use a Kalman filter to track the state of the timescale difference. We can also measure the frequency of the GNSS disciplined clock and compare it to the reference oscillator.
Over short periods of time, both legitimate GNSS time and onboard precision reference are stable and with negligible drift. The Kalman filter allows the estimation of the inter-clock time and frequency offset. At the update step, the system can reject new measurements based on the confidence interval derived from the covariance update step. Specifically, new measurements can be discarded by applying a test as \cref{eq:equation_test_single}, where $S,H,P,R$ are the covariance, measurements, prediction, and measurement noise matrices. $\hat{x}$ is the measured vector of offset and drift for each clock in the system, which should be within the $S$ confidence values of the predicted $x$.

\begin{equation}
    \begin{aligned}
         & S(t_{n+1}) = H(t_{n+1})P(t_{n+1|n})H^T(t_{n+1}) + R \\
         & \hat{x} = x(t_{n+1|n}) \pm diag(S(t_{n+1}))
    \end{aligned}
    \label{eq:equation_test_single}
\end{equation}

The test from \cref{eq:equation_test_single} only considers one measurement at a time. To improve the detection ratio, the intuitive extension is to consider a sequence of $m$ state estimates $x$  such as $x_m = [x(t_{n+1|n}), x(t_{n|n-1}, ... x(t_{n+1-m|n-m}))$ on which the windowed statistic from \cref{eq:window_stat} is defined, where $\sigma_{x_m}^2, E[x_m]$ are the estimate variance and expected values within the window.

\begin{equation}
    p(x_m) = \frac{1}{\sqrt(2\pi\sigma_{x_m}^2)}exp(-\frac{E[x_m]}{\sigma_{x_m}^2})
    \label{eq:window_stat}
\end{equation}

From the distribution in \cref{eq:window_stat}, the smoothed log-likelihood with smoothing factor, $\alpha$, in \cref{eq:smooth_log_like} is a viable quantitative test, whose responsiveness can be changed by tuning the smoothing parameter $\alpha$ to increase the robustness to measurement noise.

\begin{equation}
    Z_{x_m, i} = \alpha(Z_{x_m, i-1}) + (1-\alpha)ln(p(x_m))
    \label{eq:smooth_log_like}
\end{equation}

Based on the smoothed log-likelihood, the test in \cref{eq:smooth_log_like_test} can be formulated, where $\lambda_{T}$ is an application-specific constant, chosen to obtain a reasonable alarm rate.

\begin{equation}
    \mathcal{H}_i =
    \begin{cases}
        \mathcal{H}_0\ \mathrm{if}\ Z_{x_m, i} < \lambda_{T} \\
        \mathcal{H}_1\ \mathrm{otherwise}
    \end{cases}
    \label{eq:smooth_log_like_test}
\end{equation}

\subsection{Integrating methods} 
\label{sec:combining}
Each component targets the problem of PNT validation from a different perspective. \cref{tab:methods_summary} summarizes the key aspects of each method. The combination is achieved based on the following parameters: availability, time assurance, time accuracy, and receiver state.
\begin{table*}[]
    \centering
        \caption{GNSS validation methods summary: feature and capability comparison.}

    \begin{tabular}{|l|c|p{20mm}|p{20mm}|p{25mm}|p{20mm}|p{20mm}|}
        \hline
        Method                      & Real-time & Cryptographic features & Accuracy                                        & Connection            & Protection features              & Continuous monitoring      \\
        \hline
        \hline
        Roughtime                   & No        & YES                    & $>100ms$                                        & YES, Once             & Only PNT jumps                   & NO (Inefficient)           \\
        \hline
        Network Time Security (NTS) & YES       & YES                    & $\approx 10\mu s$                               & YES, Sparse           & Limited drag detection           & YES (Connectivity limited) \\
        \hline
        Embedded oscillator         & YES       & N/A                    & $<10ns$, within the oscillator stability window & NO (only calibration) & Lift-off and drag, no cold start & YES                        \\
        \hline
    \end{tabular}
    \vspace{4mm}
    \label{tab:methods_summary}
\end{table*}

The initial PNT solution provided by the GNSS receiver is tested against any available network-based time provider. Once a validated initial PNT is available (based, e.g. on Roughtime), the onboard oscillator and the NTS server pool can be combined to provide continuous tracking and monitoring of the PNT solution, based on the available connectivity. In particular, the NTS infrastructure is used to periodically tune the local oscillator, which, due to the high stability, allows for a longer polling time of the NTS server, preserving bandwidth and computational resources. This approach allows the defense mechanism to survive even prolonged network outages, still being able to validate the GNSS time progress


In case of a loss of lock, the system needs to take into account the duration of the outage: temporary and short-lived outages are recovered based on the embedded clock status and (depending on the connectivity availability) the NTS server pool, but prolonged outages (particularly when the adversary also limits the access to the network) require a reset of the receiver to cold start. Intuitively, long outages are comparable to situations of degraded knowledge of the constellation status, hence a cold start. The duration of the outage can be application-specific, but should not be longer than the validity of the satellite ephemeris.
		
	\section{Implementation}
	\label{section:implementation}
		
\begin{figure*}
		\centering 
		\includegraphics[width=0.95\textwidth]{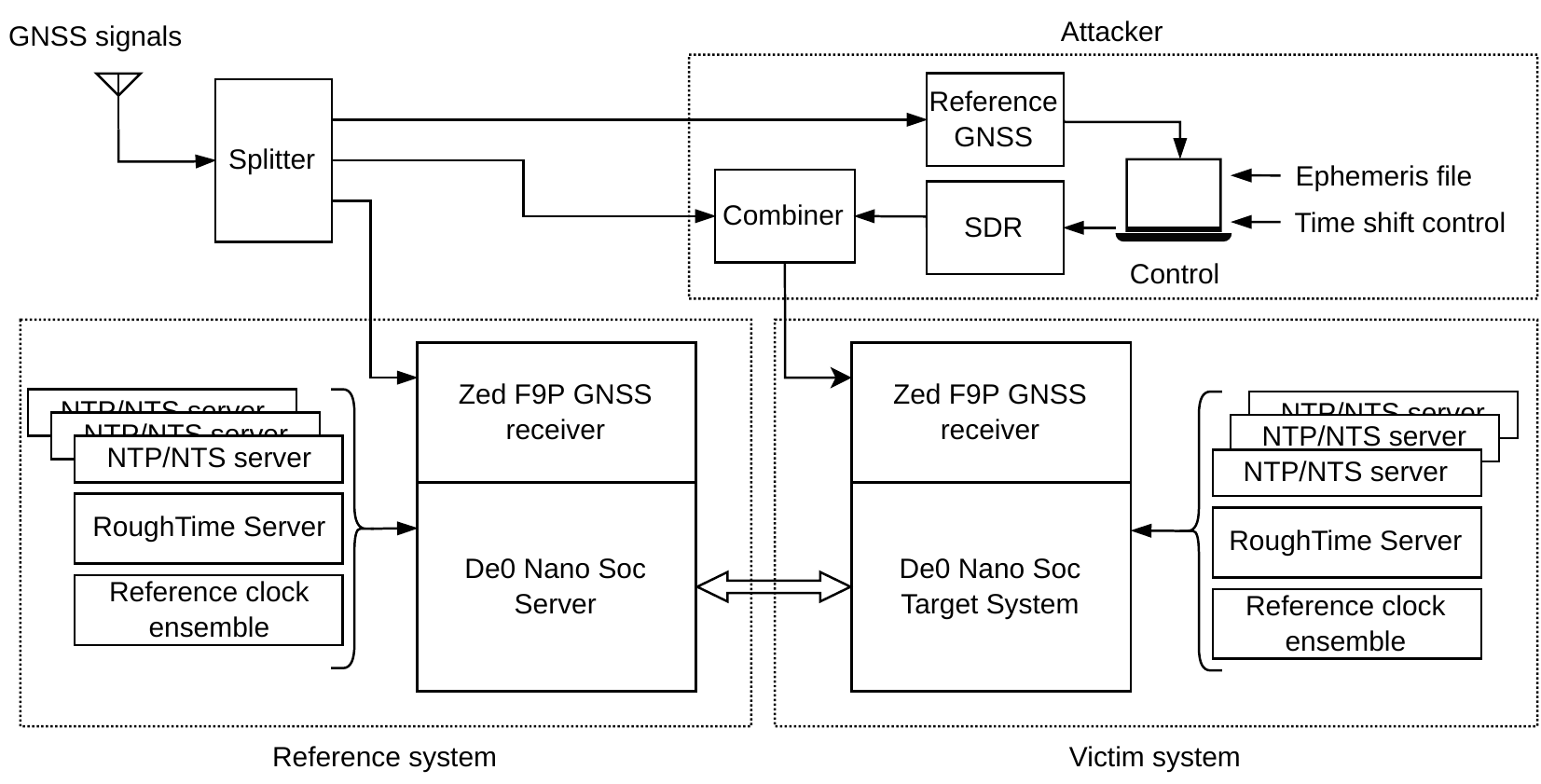}
		\caption{Experimental testbed for time-based GNSS validation.}
		\label{fig:testbed-schema}
	\end{figure*}

Experimental validation is provided by the implementation of a proof of concept capable of recreating different attack scenarios (signal simulation, signal lift-off, and replay attacks) targeting a system implementing the components described in \cref{section:methodology}. \cref{fig:testbed-schema} shows the structure of the testbed and the different components. The detection system is evaluated using publicly available network time infrastructure and a dedicated clock ensemble. Two nodes are deployed based on an Altera DE0-SOC FPGA each provided with a ZED-F9P GNSS receiver used to obtain the PNT solution. One node is used as a reference system to validate the results. The second node is the experiment target, connected to the GNSS spoofer. Additionally, each device is provided with a precision embedded clock. 

Both nodes are provided with internet connectivity and the same set of remote time servers. For monitoring purposes of the experimental testbed, the target system can measure its local clock offset against the reference system directly, on a local network connection. An attacker capable of spoofing GPS signals is connected to the victim system, which receives a combination of original and spoofed signals. The spoofer is capable of deploying both asynchronous simulation scenarios or synchronizing the spoofing signals with the real constellation using the reference receiver. Due to regulatory limitations, the transmission of the spoofing signals is conducted over cable, and the attacker-simulated signals are combined with the original signals from the antenna at the victim receiver.  
		
\section{Results and Analysis}
	\label{section:results-conclusions}
		
\begin{figure}
		\centering 
		\includegraphics[width=\linewidth]{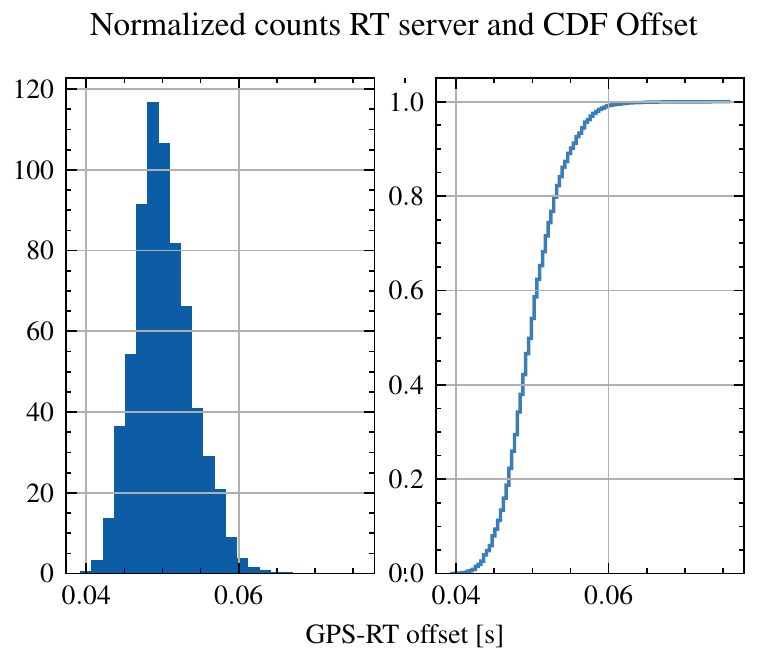}
		\caption{Roughtime server test based on available publicly available Cloudflare service. Normalized counts of RT server (left) and Cumulative Distribution Function of the offset (right).}
		\label{fig:roughtime-stable}
	\end{figure}

	The presented method proved capable of detecting the presence of an adversary, correctly identifying attacks causing lift-off of the time solution and attacks causing step changes in the GNSS PNT. Coarse simulation GNSS attackers, not performing any synchronization to the GPS frame, cause the receiver time solution to skip up to \SI{30}{\second}. Due to the lack of frame synchronization, the adversary starts transmitting the current frame with an arbitrary time offset between the spoofer timescale and the real GPS timescale.
	The Roughtime protocol specification provides a time radius for publicly available Roughtime servers as wide as \SI{10}{s}, which is enough to provide spoofing detection against simulation adversaries. On the other hand, an evaluation of the publicly available Roughtime server implementations shows that better accuracy is possible, specifically when in conjunction with NTS. Testing the \textit{cloudflare.roughtime.se} public server against a GNSS-disciplined clock, shows that the stability and accuracy of the Roughtime timestamp are better than the values in the standard definition as shown in \cref{fig:roughtime-stable}. Specifically, the Cloudflare server declared radius is \SI{1}{\second}, which is consistent with measurements obtained from the GNSS receiver. The specific method used to discipline the clock of the remote time server, used to produce the Roughtime timestamps, is unknown.

\begin{figure}
		\centering 
		\includegraphics[width=\linewidth]{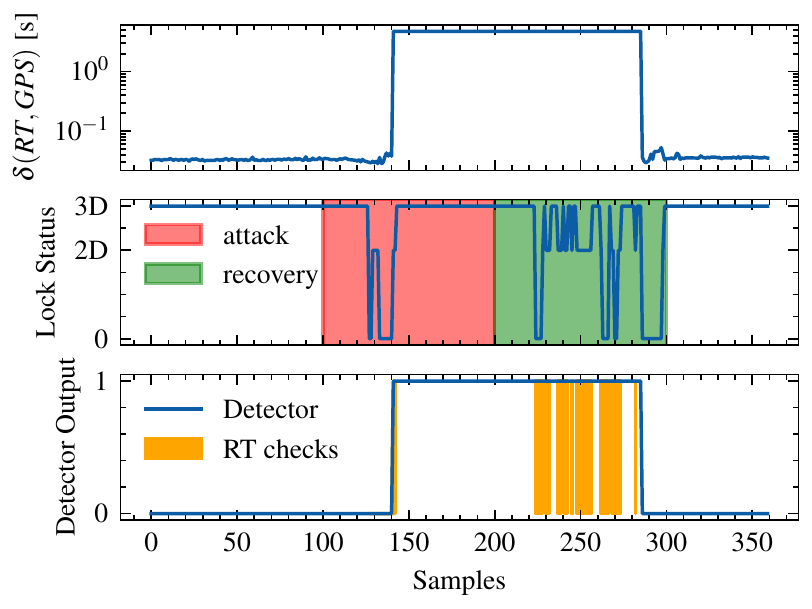}
		\caption{Roughtime based detector: detection of step change GPS time offset.}
		\label{fig:roughtime-detector-step}
	\end{figure}

    \cref{fig:roughtime-detector-step} shows the detection of a coarse adversary based on simulation not synchronized with the original signals. A GNSS receiver is given enough time to obtain a clean PNT solution (time step up to sample 100). At time mark 100, the simulator is enabled. The simulation is configured to be consistent with the current constellation regarding the number of satellites and satellite position, but the attacker does not attempt any alignment of the simulation time with the real constellation time. As a matter of fact, this translated into the attacker causing a sharp discontinuity at the beginning and the end of the simulation. The output of the detector proposed in \cref{eq:hypothesis-rt} is shown in \cref{fig:roughtime-detector-step}, where an attacker causes an offset of \SI{4}{\second} between the real GNSS time and the simulated one. 
 
\begin{figure}
		\centering 
		\includegraphics[width=\linewidth]{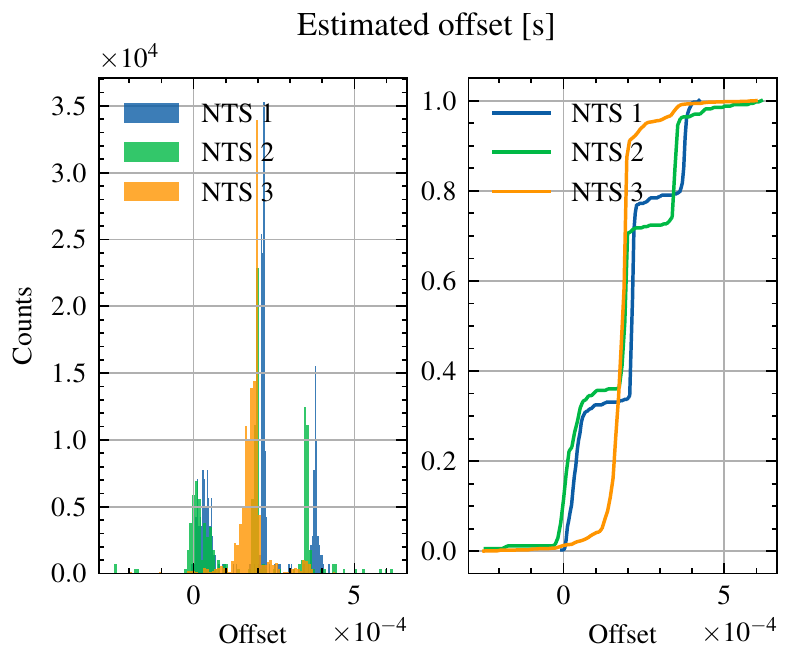}
		\caption{NTS stability analysis and detection threshold estimation, three remote peers located in geographical proximity to the test device are used. Normalized counts (left) and Cumulative Distribution Function of the offset (right).}
		\label{fig:nts-stable}
	\end{figure}

 As discussed in \cref{section:sys-adversary-model}, attackers capable of more advanced synchronization of the simulated signals (i.e., the transmission time is aligned with the beginning of the frame) will cause a smaller time step during the initial takeover, requiring higher detection accuracy. NTS allows network-connected devices to achieve a synchronization offset with remote secure time servers as low as \SI{150}{\micro\second}, with a standard deviation of \SI{50}{\micro\second}, depending on the relative location of the server and client and the stability of the network connectivity. \cref{fig:nts-stable} shows normalized counts and cumulative probability functions for four independent NTS servers, obtained over long observations over a stable network.    
		
\begin{figure}
		\centering 
		\includegraphics[width=\linewidth]{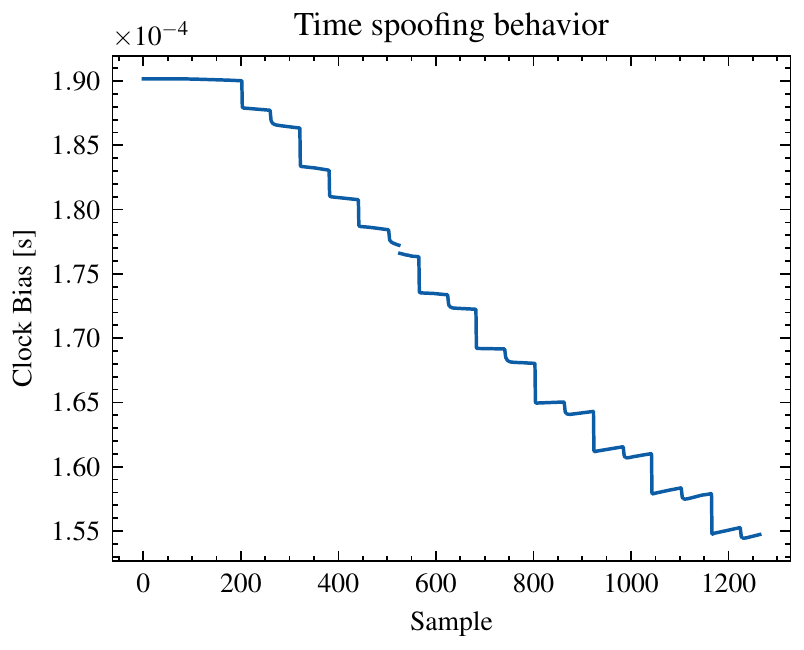}
		\caption{Receiver clock bias under attack. Incremental progression of the GNSS clock bias can be seen and it is consistent with the attacker profile.}
		\label{fig:ublox-offset}
	\end{figure}

To circumvent tests based on Roughtime, the attacker needs to align the beginning of the GNSS frame to the real transmission time. This is achieved by synchronizing the radio transmitter front-end to the spoofing target constellation. The spoofer then modifies the content of the spoofed navigation message to force the receiver to re-compute its clock offset, progressively deviating it from the GNSS timescale. Tests performed with the testbed from \cref{fig:testbed-schema} and real GPS signals show that it is possible to capture the Ublox F9P receiver. 
Notably, spoofing corrections of the PNT time solution are applied coherently with the start of the frame. At these times, the receiver does not produce a PNT solution until the next subframe is received. Additionally, no loss of tracking was noticed, which makes us believe the internal PNT engine implemented by Ublox keeps into account changes in the PNT solution but eventually accepts the changes nonetheless. This can be seen in \cref{fig:ublox-offset}, where periodically \SI{2}{\micro\second} are added to the GNSS time solution, progressively shifting it. 
		
\begin{figure}
		\centering 
		\includegraphics[width=\linewidth]{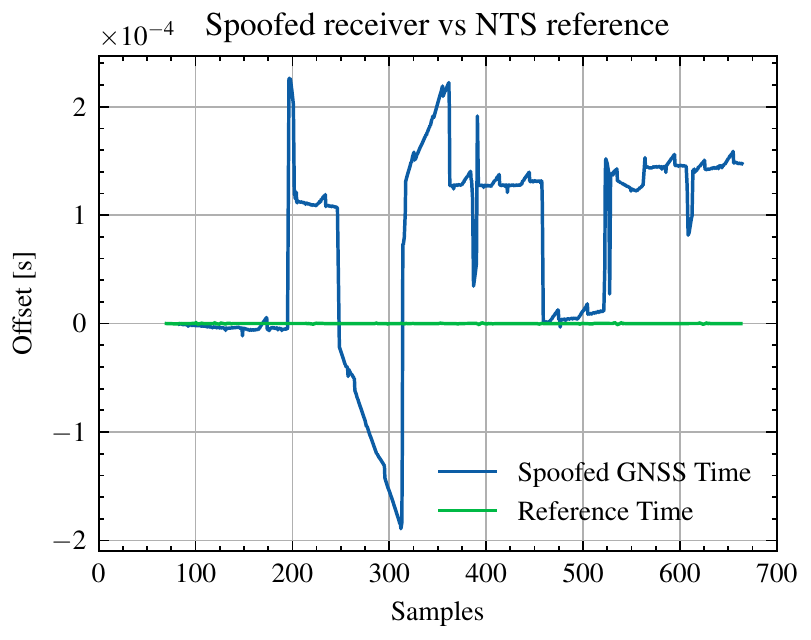}
		\caption{Receiver clock offset under attack, detection based on comparison of remote time servers.}
		\label{fig:victim-receiver-offset}
	\end{figure}

The test implemented according to \cref{eq:hypothesis-nts} is capable of detecting the spoofing attack based on the time difference between the reference NTS source and the GNSS time information. Clear discrepancies caused by the attack signals are visible in \cref{fig:victim-receiver-offset}. The detection threshold is determined based on the confidence interval of the NTS time offset, calculated from the standard deviation of long-term observations of the remote time server. 

\begin{figure}
		\centering 
		\includegraphics[width=\linewidth]{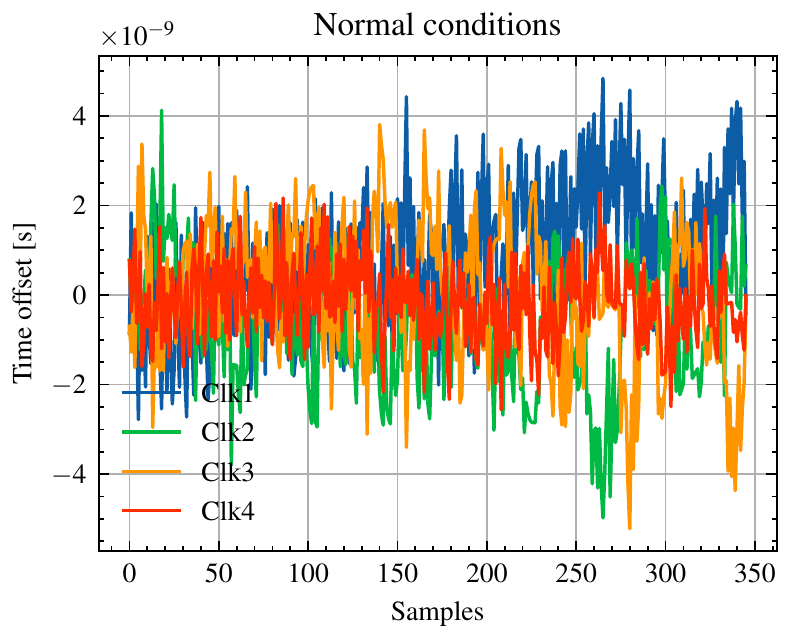}
		\caption{Comparison of the un-spoofed PNT time solution of the GNSS receiver with local clock ensemble: the average error is contained with \SI{10}{\nano\second}.}
		\label{fig:clock_reference-offset}
	\end{figure}

\begin{figure}
		\centering 
		\includegraphics[width=\linewidth]{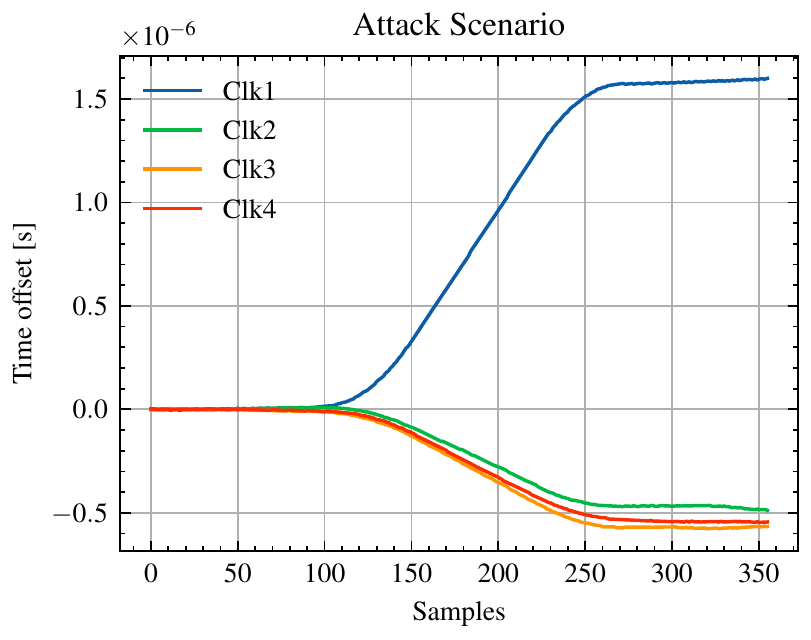}
		\caption{Comparison of the spoofed PNT time solution of the GNSS receiver with local clock ensemble, showing discrepancies with the local ensemble time progression.}
		\label{fig:clock_attack-offset}
	\end{figure}

\begin{figure}
		\centering 
		\includegraphics[width=\linewidth]{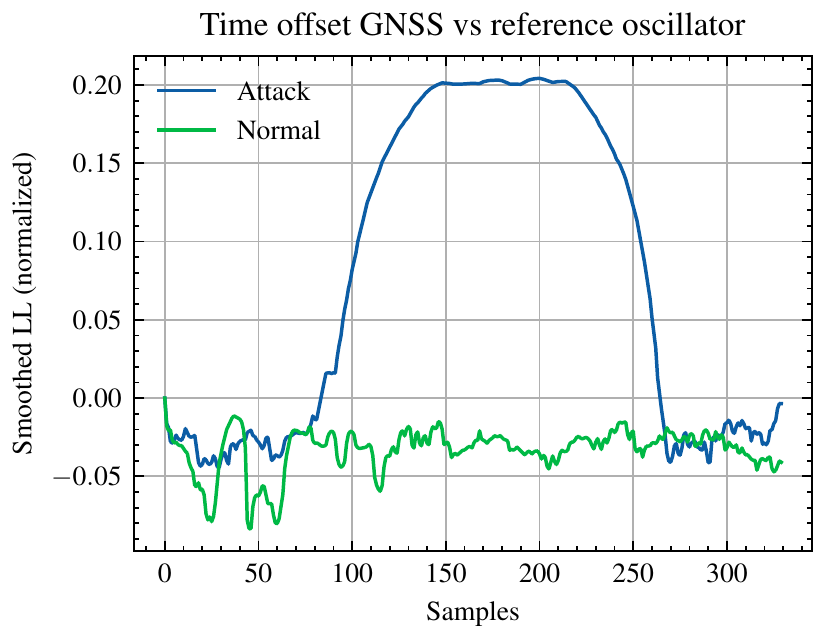}
		\caption{Result of the smoothed negative log-likelihood test for the spoofed and un-spoofed case.}
		\label{fig:detector_ll}
	\end{figure}

Ultimately, better synchronization of the simulated signals, taking into account the exact code phase and carrier phase and Doppler shift at the victim antenna, allows the adversary to produce seamless lift-off of the time solution at the victim receiver. In this case, the initial time offset of the GNSS receiver is below the detection threshold for the NTS-based test. Similarly, the same happens when the attacker's target is to produce a total time variation that falls within the NTS reference server standard deviation. In this case, a precision local oscillator (or a multiplicity of them) is required to compare the GNSS time solution's short-term trend and stability. The time offset between benign GNSS solution and an ensemble of reference clocks is within \SI{10}{\nano\second} (\cref{fig:clock_reference-offset}). The attack consists of a \SI{2}{\micro\second} smooth time pull which is visible and correctly detected by a corresponding deviation in the clock offset between the GNSS receiver and the local clock ensemble, shown in \cref{fig:clock_attack-offset}. The detector proposed in \cref{eq:smooth_log_like_test} detects the attack (\cref{fig:detector_ll}) as expected. 

Notably, the long-term stability of local oscillators depends on the quality of the oscillator itself. For this reason, periodic re-synchronization of the local oscillator is required and this can be performed using a trusted NTS source. Nevertheless, more accurate embedded oscillators allow for progressively longer periods where connectivity is not guaranteed. 

The validation of the PNT solution can be potentially performed at each PNT update. The low computation cost of the local validation makes this possible, as long as the platform is appropriately provisioned. On the other hand, resorting to external references is subject to networking and other processing delays, including cryptographic overhead (\cref{section:cryptocost}). 
A typical mode of operation relies on continuous local validation interleaved with less frequent remote validation.

\subsection{Cryptographic overhead considerations}	
\label{section:cryptocost}
Roughtime implements public key cryptography, notably digital signatures based on Ed25519 Edwards-curve Digital Signature Algorithm (EcDSA). 
The overhead can be negligible on powerful consumer machines, but it can be significant on constrained embedded platforms (as those that typically combined with embedded GNSS receivers). \cref{tab:RT_cost} compares three CPUs of different tiers (embedded ARM, laptop and desktop), oftware-based implementation (OpenSSL). The architecture used by the DE0 Nano SOC embedded ARM is Cortex A9 armv7l (without hardware cryptographic accelerators) and the system requires \SI{1.5}{\milli\second} per signature verification, showing it is still possible to deploy Roughtime on constrained embedded systems. Cryptographic co-processors, increasingly diffused even on lower-tier CPUs, can improve the signature verification rate. 
	
	\begin{table*}[th!]
		\centering
  		\caption{Roughtime message verification time.}

		\begin{tabular}{@{}p{35mm}p{15mm}p{15mm}rrrrrr@{}}
			
			CPU                         & Architecture & Frequency              & sign (Digest)                      & verify (Digest)                   & sign/s  & verify/s \\
			\hline
			Cortex A9                   & armv7l       & \SI{800}{\mega\hertz}  & \SI{6e-1}{\milli\second}   & \SI{1.5}{\milli\second}   & 1786.6  & 666.7    \\
			Intel(R) Core(TM) i7-8750H  & x86          & \SI{2200}{\mega\hertz} & \SI{4.41e-2}{\milli\second} & \SI{1.09e-1}{\milli\second} & 24303.9 & 9134.3   \\
			Intel(R) Xeon(R) E5-1660 v3 & x86          & \SI{3000}{\mega\hertz} & \SI{4.46e-2}{\milli\second} & \SI{1.24e-1}{\milli\second} & 21737.5 & 8044.1   \\
		\end{tabular}
		\label{tab:RT_cost}
	\end{table*}

	\begin{table*}[th!]
		\centering
  		\caption{AEAD\_AES\_SIV\_CMAC\_256 performance comparison (results are in \si{\operations/\second}).}

		\begin{tabular}{@{}p{35mm}p{15mm}p{15mm}rrrrrr@{}}
			CPU                         & Architecture & Frequency              &  Average Latency & Block 1024 bytes & Block 8192 bytes  \\
			\hline
			Cortex A9                   & armv7l       & \SI{800}{\mega\hertz}  & \SI{4.52e-5}{\milli\second}   & 21.25M     & 22.96M \\
			Intel(R) Core(TM) i7-8750H  & x86          & \SI{2200}{\mega\hertz} & \SI{3.30e-7}{\milli\second}   & 2211.69M   & 3842.31M  \\
			Intel(R) Xeon(R) E5-1660 v3 & x86          & \SI{3000}{\mega\hertz} & \SI{4.65e-7}{\milli\second}   & 1812.40M   & 2487.03M \\
		\end{tabular}

		\label{tab:NTS_cost}
	\end{table*}
 
Compared to Roughtime, NTS deploys a more lightweight cryptographic scheme that consists of a one-time key establishment, and symmetric key cryptography operations.  \cref{tab:NTS_cost} (for the same platforms and software implementation) demonstrates that for different packet sizes (NTS packet size: 228 bytes to 1468 bytes) AES256\_AEAD\_CMAC used by NTS introduces a negligible overhead, allowing continuous operation based on connection availability (Note: network latency not considered here). 

\section{Conclusion and Future Work}
\label{section:conclusion}
We present three different components for GNSS time-based validation that, combined, provide all-around coverage, nullifying the possibility of undetected (asynchronous or synchronous) simulation-based attacks modifying the GNSS receiver time solution. Reduced cryptographic overhead by leveraging different secure remote time providers and availability of offline reference sources allow continuous monitoring of the time part of the PNT solution. The proposed system is an adaptable countermeasure that can easily be deployed to protect existing receivers from coarse and advanced spoofing attacks. On the other hand, adversaries capable of modifying the PN part of the GNSS solution, without modifying the time solution cannot be detected by countermeasures based on time validation. This is true for the method presented here or any other time-based countermeasure. To the best of the authors' knowledge, a trauma on the time offset is often present at take-over stage, even if the latter is smooth. Further investigation regarding attacks that do not produce phase and frequency offsets at the initial stage is in progress.
Nevertheless, this method is capable of efficient and secure time-based verification, improving the existing literature and leveraging the public secure time infrastructure in combination with local precision clocks. Future analysis regarding the inclusion of potentially malicious sources is required. In particular, the case of authenticated time providers colluding with a GNSS attacker requires further investigation, and a different rejection method is being explored, which takes into account multiple remote time references of varying security levels. 

\section*{Acknowledgment}
This work was supported in part by the SSF SURPRISE cybersecurity project and the Security Link strategic research center.

\bibliographystyle{ieeetran}
\bibliography{refactorbib}

\begin{thebibliography}{10}
\providecommand{\url}[1]{#1}
\csname url@samestyle\endcsname
\providecommand{\newblock}{\relax}
\providecommand{\bibinfo}[2]{#2}
\providecommand{\BIBentrySTDinterwordspacing}{\spaceskip=0pt\relax}
\providecommand{\BIBentryALTinterwordstretchfactor}{4}
\providecommand{\BIBentryALTinterwordspacing}{\spaceskip=\fontdimen2\font plus
\BIBentryALTinterwordstretchfactor\fontdimen3\font minus
  \fontdimen4\font\relax}
\providecommand{\BIBforeignlanguage}[2]{{%
\expandafter\ifx\csname l@#1\endcsname\relax
\typeout{** WARNING: IEEEtran.bst: No hyphenation pattern has been}%
\typeout{** loaded for the language `#1'. Using the pattern for}%
\typeout{** the default language instead.}%
\else
\language=\csname l@#1\endcsname
\fi
#2}}
\providecommand{\BIBdecl}{\relax}
\BIBdecl

\bibitem{Thombre2018}
S.~Thombre, M.~Z.~H. Bhuiyan, P.~Eliardsson \emph{et~al.}, ``{GNSS} threat
  monitoring and reporting: Past, present, and a proposed future,''
  \emph{Navigation, Journal of the Institute of Navigation}, vol.~71, no.~3,
  pp. 513--529, 2018.

\bibitem{psiaki2016gnss}
M.~L. Psiaki and T.~E. Humphreys, ``{GNSS} spoofing and detection,''
  \emph{Proceedings of the IEEE}, vol. 104, no.~6, pp. 1258--1270, 2016.

\bibitem{Fernandez-Hernandez2016}
I.~Fern{\'{a}}ndez-Hern{\'{a}}ndez, V.~Rijmen, G.~Seco-Granados \emph{et~al.},
  ``A navigation message authentication proposal for the galileo open
  service,'' \emph{Navigation, Journal of the Institute of Navigation},
  vol.~63, no.~1, pp. 85--102, 3 2016.

\bibitem{anderson2017chips}
J.~M. Anderson, K.~L. Carroll, N.~P. DeVilbiss \emph{et~al.}, ``Chips-message
  robust authentication (chimera) for gps civilian signals,'' in \emph{30th
  International Technical Meeting of the Satellite Division of the Institute of
  Navigation, ION {GNSS}+ 2017}, Portland, Oregon, 2017, pp. 2388--2416.

\bibitem{Gamba2021ComputationalPlatforms}
M.~T. Gamba, M.~Nicola, and B.~Motella, ``Computational load analysis of a
  galileo osnma-ready receiver for arm-based embedded platforms,''
  \emph{Sensors}, vol.~21, no.~2, pp. 1--21, 1 2021.

\bibitem{Cucchi2021AssessingReceiver}
L.~Cucchi, S.~Damy, M.~Paonni \emph{et~al.}, ``Assessing galileo osnma under
  different user environments by means of a multi-purpose test bench, including
  a software-defined {GNSS} receiver,'' in \emph{34th International Technical
  Meeting of the Satellite Division of the Institute of Navigation, ION {GNSS}+
  2021}, St. Louis, Missouri, 2021, pp. 3653--3667.

\bibitem{Motella2020AReceiver}
B.~Motella, M.~T. Gamba, and M.~Nicola, ``A real-time osnma-ready software
  receiver,'' in \emph{International Technical Meeting of The Institute of
  Navigation (ITM 2020)}, San Diego, California, 2020, pp. 979--991.

\bibitem{Gotzelmann2021GalileoProvision}
M.~G{\"{o}}tzelmann, E.~K{\"{o}}ller, I.~V. Semper \emph{et~al.}, ``Galileo
  open service navigation message authentication: Preparation phase and drivers
  for future service provision,'' in \emph{34th International Technical Meeting
  of the Satellite Division of the Institute of Navigation, ION {GNSS}+ 2021},
  St. Louis, Missouri, 2021, pp. 385--401.

\bibitem{Lenhart2022}
M.~Lenhart, M.~Spanghero, and P.~Papadimitratos, ``{Distributed and Mobile
  Message Level Relaying/Replaying of GNSS Signals},'' in \emph{2022
  International Technical Meeting of The Institute of Navigation (ITM)}, Long
  Beach, California, 2022, pp. 56--57.

\bibitem{Seco-Granados2021}
G.~Seco-Granados, D.~G{\'{o}}mez-Casco, J.~A. L{\'{o}}pez-Salcedo
  \emph{et~al.}, ``Detection of replay attacks to {GNSS} based on partial
  correlations and authentication data unpredictability,'' \emph{GPS
  Solutions}, vol.~25, no.~2, 2021.

\bibitem{papadimMilcom2008}
P.~Papadimitratos and A.~Jovanovic, ``{GNSS-based Positioning: Attacks and
  Countermeasures},'' in \emph{IEEE Military Communications Conference (IEEE
  MILCOM)}, San Diego, CA, USA, 2008, pp. 1--7.

\bibitem{jafarnia2013PNT}
A.~Jafarnia-Jahromi, S.~Daneshmand, A.~Broumandan \emph{et~al.}, ``Pvt solution
  authentication based on monitoring the clock state for a moving {GNSS}
  receiver,'' in \emph{European Navigation Conference (ENC2013)}, Vienna,
  Austria, 2013.

\bibitem{Arafin2016DetectingOscillators}
M.~T. Arafin, D.~M. Anand, and G.~Qu, ``Detecting {GNSS} spoofing using a
  network of hardware oscillators,'' in \emph{Annual Precise Time and Time
  Interval Systems and Applications Meeting (PTTI)}, Monterey, California,
  2016, pp. 74--79.

\bibitem{Hwang2014}
P.~Y. Hwang and G.~A. McGraw, ``Receiver autonomous signal authentication
  (rasa) based on clock stability analysis,'' in \emph{IEEE/ION Position,
  Location and Navigation Symposium (PLANS 2014)}, Monterey, California, 2014,
  pp. 270--281.

\bibitem{Spanghero2022}
M.~Spanghero and P.~Papadimitratos, ``{High-precision Hardware Oscillators
  Ensemble for GNSS Attack Detection},'' in \emph{2022 IEEE Aerospace
  Conference}, Big Sky, Montana, 2022, pp. 1--11.

\bibitem{spangheroGNSS20}
M.~Spanghero, K.~Zhang, and P.~Papadimitratos, ``{Authenticated time for
  detecting GNSS attacks},'' in \emph{33rd International Technical Meeting of
  the Satellite Division of the Institute of Navigation (ION GNSS+)}, Online,
  virtual, 2020, pp. 3826--3834.

\bibitem{kzmsppPLANS2020}
K.~Zhang, M.~Spanghero, and P.~Papadimitratos, ``{Protecting GNSS-based
  Services using Time Offset Validation},'' in \emph{IEEE/ION Position,
  Location and Navigation Symposium (PLANS)}, Portland, Oregon, 2020, pp.
  575--583.

\bibitem{tippenhauer2011requirements}
N.~O. Tippenhauer, C.~P\"{o}pper, K.~B. Rasmussen \emph{et~al.}, ``On the
  requirements for successful gps spoofing attacks,'' in \emph{18th ACM
  Conference on Computer and Communications Security (ACM CCS)}, New York, NY,
  USA, 2011, p. 75–86.

\bibitem{Kerns2014}
A.~J. Kerns, D.~P. Shepard, J.~A. Bhatti \emph{et~al.}, ``Unmanned aircraft
  capture and control via {GPS} spoofing,'' \emph{Journal of Field Robotics},
  vol.~31, no.~4, pp. 617--636, Apr. 2014.

\bibitem{Bhatti2017}
J.~Bhatti and T.~E. Humphreys, ``Hostile control of ships via false {GPS}
  signals: Demonstration and detection,'' \emph{Navigation, Journal of the
  Institute of Navigation}, vol.~64, no.~1, pp. 51--66, Mar. 2017.

\bibitem{Ioannides2016}
R.~T. Ioannides, T.~Pany, and G.~Gibbons, ``Known vulnerabilities of global
  navigation satellite systems, status, and potential mitigation techniques,''
  \emph{Proceedings of the {IEEE}}, vol. 104, no.~6, pp. 1174--1194, Jun. 2016.

\bibitem{KexiongAllBelongToUs2018}
\BIBentryALTinterwordspacing
K.~C. Zeng, S.~Liu, Y.~Shu \emph{et~al.}, ``All your {GPS} are belong to us:
  Towards stealthy manipulation of road navigation systems,'' in \emph{27th
  USENIX Security Symposium (USENIX Security 18)}, Baltimore, MD, 2018, pp.
  1527--1544. [Online]. Available:
  \url{https://www.usenix.org/conference/usenixsecurity18/presentation/zeng}
\BIBentrySTDinterwordspacing

\bibitem{Feng2021}
W.~Feng, J.-M. Friedt, G.~Goavec-Merou \emph{et~al.}, ``Software-defined radio
  implemented {GPS} spoofing and its computationally efficient detection and
  suppression,'' \emph{{IEEE} Aerospace and Electronic Systems Magazine},
  vol.~36, no.~3, pp. 36--52, 2021.

\bibitem{HuangL2015}
L.~Huang and Q.~Yang, ``Low-cost gps simulator - gps spoofing by sdr,'' in
  \emph{Proceedings of DEF CON23}, Las Vegas, NV , USA, 2015.

\bibitem{LeksellTGalileo2021}
\BIBentryALTinterwordspacing
T.~Leksell, ``A comparison of smartphone gpsl1 and galileo e1-b/c spoofing
  resilience,'' Master's thesis, KTH, School of Electrical Engineering and
  Computer Science (EECS), 2021. [Online]. Available:
  \url{http://kth.diva-portal.org/smash/record.jsf?pid=diva2:1544978}
\BIBentrySTDinterwordspacing

\bibitem{SDRMultiFrequency2018}
\BIBentryALTinterwordspacing
``{(In)Feasibility of Multi-Frequency Spoofing},'' Inside {GNSS} - Global
  Navigation Satellite Systems Engineering, Policy, and Design, June 2018.
  [Online]. Available:
  \url{https://insidegnss.com/infeasibility-of-multi-frequency-spoofing/}
\BIBentrySTDinterwordspacing

\bibitem{HumphreysAssessingSpoofer}
T.~E. Humphreys, B.~M. Ledvina, M.~L. Psiaki \emph{et~al.}, ``Assessing the
  spoofing threat: Development of a portable gps civilian spoofer,'' in
  \emph{21st International Technical Meeting of the Satellite Division of The
  Institute of Navigation (ION GNSS 2008)}, Savannah, Georgia, 1987, pp.
  2314--2325.

\bibitem{Maier2018}
D.~S. Maier, K.~Frankl, and T.~Pany, ``The {{GNSS}}-transceiver: Using
  vector-tracking approach to convert a {{GNSS}} receiver to a simulator:
  Implementation and verification for signal authentication,'' in \emph{31st
  International Technical Meeting of The Satellite Division of the Institute of
  Navigation ({ION} {{GNSS}} 2018)}, Miami, Florida, 2018, pp. 4231--4244.

\bibitem{Shepard2012c}
D.~P. Shepard, T.~E. Humphreys, and A.~A. Fansler, ``Evaluation of the
  vulnerability of phasor measurement units to gps spoofing attacks,''
  \emph{International Journal of Critical Infrastructure Protection}, vol.~5,
  no. 3-4, pp. 146--153, 2012.

\bibitem{Humphreys2012}
T.~Humphreys, J.~Bhatti, D.~Shepard \emph{et~al.}, ``The texas spoofing test
  battery: Toward a standard for evaluating gps signal authentication
  techniques,'' in \emph{25th International Technical Meeting of the Satellite
  Division of the Institute of Navigation 2012, ION {GNSS}+}, Nashville,
  Tennessee, 2012, pp. 3569--3583.

\bibitem{Jiang2013}
X.~Jiang, J.~Zhang, B.~J. Harding \emph{et~al.}, ``Spoofing {GPS} receiver
  clock offset of phasor measurement units,'' \emph{{IEEE} Transactions on
  Power Systems}, vol.~28, no.~3, pp. 3253--3262, Aug. 2013.

\bibitem{Zhu2016}
F.~Zhu, A.~Youssef, and W.~Hamouda, ``Detection techniques for data-level
  spoofing in {GPS}-based phasor measurement units,'' in \emph{2016
  International Conference on Selected Topics in Mobile and Wireless Networking
  ({MoWNeT})}, Cairo, Egypt, 2016, pp. 1--8.

\bibitem{Zhang2013}
Z.~Zhang, S.~Gong, A.~D. Dimitrovski \emph{et~al.}, ``Time synchronization
  attack in smart grid: Impact and analysis,'' \emph{{IEEE} Transactions on
  Smart Grid}, vol.~4, no.~1, pp. 87--98, Mar. 2013.

\bibitem{humphreys2013detection}
T.~E. Humphreys, ``Detection strategy for cryptographic {GNSS} anti-spoofing,''
  \emph{IEEE Transactions on Aerospace and Electronic Systems}, vol.~49, no.~2,
  pp. 1073--1090, 2013.

\bibitem{Arizabaleta2019}
M.~Arizabaleta, E.~Gkougkas, and T.~Pany, ``A feasibility study and risk
  assessment of security code estimation and replay ({SCER}) attacks,'' in
  \emph{32nd International Technical Meeting of the Satellite Division of The
  Institute of Navigation ({ION} {{GNSS}+} 2019)}, Miami, Florida, 2019, pp.
  1039--1050.

\bibitem{Gallardo2020}
F.~Gallardo and A.~P. Yuste, ``{SCER} spoofing attacks on the galileo open
  service and machine learning techniques for end-user protection,''
  \emph{{IEEE} Access}, vol.~8, pp. 85\,515--85\,532, 2020.

\bibitem{Akos2012}
D.~M. Akos, ``Who's afraid of the spoofer? gps/{GNSS} spoofing detection via
  automatic gain control (agc),'' \emph{Navigation, Journal of the Institute of
  Navigation}, vol.~59, no.~4, pp. 281--290, 12 2012.

\bibitem{Lo2019}
S.~Lo, Y.~H. Chen, D.~Akos \emph{et~al.}, ``Test of crowdsourced smartphones
  measurements to detect {{GNSS}} spoofing and other disruptions,'' in
  \emph{The International Technical Meeting of the The Institute of
  Navigation}, Reston, Virginia, 2019, pp. 373--388.

\bibitem{wesson2017gnss}
K.~D. Wesson, J.~N. Gross, T.~E. Humphreys \emph{et~al.}, ``{GNSS} signal
  authentication via power and distortion monitoring,'' \emph{IEEE Transactions
  on Aerospace and Electronic Systems}, vol.~54, no.~2, pp. 739--754, 2018.

\bibitem{psiaki2013antenna}
M.~L. Psiaki, S.~P. Powell, and B.~W. O'Hanlon, ``{GNSS} spoofing detection
  using high-frequency antenna motion and carrier-phase data,'' in \emph{26th
  International Technical Meeting of the Satellite Division of the Institute of
  Navigation, ION {GNSS}+ 2013}, vol.~4, Nashville, Tennessee, 2013, pp.
  2949--2991.

\bibitem{Jada2021}
S.~Jada, M.~Psiaki, S.~Landerkin \emph{et~al.}, ``Evaluation of {PNT}
  situational awareness algorithms and methods,'' in \emph{34th International
  Technical Meeting of the Satellite Division of The Institute of Navigation
  ({ION} {{GNSS}+} 2021)}, St. Louis, Missouri, 2021, pp. 816--833.

\bibitem{Sathaye2020}
H.~Sathaye, G.~LaMountain, P.~Closas \emph{et~al.}, ``Semperfi: Anti-spoofing
  {GPS} receiver for uavs,'' in \emph{29th Annual Network and Distributed
  System Security Symposium (NDSS)}, San Diego, California, 2022.

\bibitem{Miralles2018}
D.~Miralles, N.~Levigne, D.~M. Akos \emph{et~al.}, ``Android raw {{GNSS}}
  measurements as the new anti-spoofing and anti-jamming solution,'' in
  \emph{31st International Technical Meeting of The Satellite Division of the
  Institute of Navigation ({ION} {{GNSS}+} 2018)}, Miami, Florida, 2018, pp.
  334--344.

\bibitem{Curran2017OnTU}
J.~T. Curran and A.~Broumandan, ``On the use of low-cost imus for {GNSS}
  spoofing detection in vehicular applications,'' in \emph{International
  Tecnincal Symposium on Navigation and Timing (ITSNT)}, Toulouse, France,
  2017.

\bibitem{Broumandan2018}
A.~Broumandan and G.~Lachapelle, ``Spoofing detection using {GNSS}/ins/odometer
  coupling for vehicular navigation,'' \emph{Sensors}, vol.~18, no.~5, 2018.

\bibitem{Clemens2022}
\BIBentryALTinterwordspacing
Z.~Clements, J.~E. Yoder, and T.~E. Humphreys, ``Carrier-phase and imu based
  gnss spoofing detection for ground vehicles,'' 2022. [Online]. Available:
  \url{https://arxiv.org/abs/2203.00140}
\BIBentrySTDinterwordspacing

\bibitem{LiuWPP:2023}
W.~Liu and P.~Papadimitratos, ``Probabilistic detection of gnss spoofing using
  opportunistic information,'' in \emph{IEEE/ION Position, Location and
  Navigation Symposium (PLANS)}, Monterey, California, 2023.

\bibitem{Arafin2017}
\BIBentryALTinterwordspacing
M.~T. Arafin, D.~Anand, and G.~Qu, ``A low-cost gps spoofing detector design
  for internet of things (iot) applications,'' in \emph{Great Lakes Symposium
  on VLSI (GLSVLSI17)}, New York, NY, USA, 2017, p. 161–166. [Online].
  Available: \url{https://doi.org/10.1145/3060403.3060455}
\BIBentrySTDinterwordspacing

\bibitem{ODriscol2020}
C.~O'Driscol, S.~Keating, and G.~Caparra, ``A performance assessment of secure
  wireless two-way time transfer,'' in \emph{33rd International Technical
  Meeting of the Satellite Division of The Institute of Navigation ({ION}
  {{GNSS}}+ 2020)}, Online, virtual, 2020, pp. 3938--3951.

\bibitem{perry2021}
Y.~Perry, N.~Rozen-Schiff, and M.~Schapira, ``A devil of a time: How vulnerable
  is ntp to malicious timeservers?'' in \emph{Proceedings 2021 Network and
  Distributed System Security Symposium}, Reston, Virginia, 2021.

\bibitem{Deutsch2018}
O.~Deutsch, N.~R. Schiff, D.~Dolev \emph{et~al.}, ``Preventing (network) time
  travel with chronos,'' in \emph{Network and Distributed System Security
  Symposium (NDSS)}, Reston, VA, 2018.

\bibitem{Malhotra2017}
A.~Malhotra, I.~E. Cohen, E.~Brakke \emph{et~al.}, ``Attacking the network time
  protocol,'' in \emph{23rd Annual Network and Distributed System Security
  Symposium (NDSS)}, San Diego, California, 2016.

\bibitem{rfc5906}
\BIBentryALTinterwordspacing
P.~D.~L. Mills and B.~Haberman, ``{Network Time Protocol Version 4: Autokey
  Specification},'' RFC 5906, Jun. 2010. [Online]. Available:
  \url{https://www.rfc-editor.org/info/rfc5906}
\BIBentrySTDinterwordspacing

\bibitem{ietf-ntp-using-nts-for-ntp-28}
D.~Franke, D.~Sibold, K.~Teichel \emph{et~al.}, ``Network time security for the
  network time protocol,'' Internet Engineering Task Force, Tech. Rep.
  draft-ietf-ntp-using-nts-for-ntp-28, 2020.

\bibitem{ietf-ntp-roughtime-07}
\BIBentryALTinterwordspacing
A.~Malhotra, A.~Langley, W.~Ladd \emph{et~al.}, ``{Roughtime},'' Sep. 2022,
  work in progress. [Online]. Available:
  \url{https://datatracker.ietf.org/doc/draft-ietf-ntp-roughtime/07/}
\BIBentrySTDinterwordspacing

\bibitem{psiaki2013}
M.~L. Psiaki, B.~W. O'Hanlon, J.~A. Bhatti \emph{et~al.}, ``Gps spoofing
  detection via dual-receiver correlation of military signals,'' \emph{IEEE
  Transactions on Aerospace and Electronic Systems}, vol.~49, no.~4, pp.
  2250--2267, 10 2013.

\end{thebibliography}
		
\end{document}